\documentclass{article}
\usepackage{arxiv}

\usepackage[utf8]{inputenc}
\usepackage[T1]{fontenc}
\usepackage{amssymb}
\usepackage{latexsym}

\usepackage{url}
\usepackage{xcolor}
\usepackage{booktabs}
\usepackage{mathrsfs}
\usepackage{amsfonts} 
\usepackage{nicefrac}       % compact symbols for 1/2, etc.
\usepackage{microtype}      % microtypography
\usepackage{lipsum}	
\usepackage{hyperref}
\usepackage{cleveref}
\usepackage{natbib}
\usepackage{latexsym}
\usepackage{graphicx}
\definecolor{newcolor}{rgb}{.8,.349,.1}
\usepackage{authblk}
\graphicspath{{./Figures/}}

\begin{document}

\title{Augmentation Based Unsupervised Domain Adaptation}%
% \tnotetext[tnote1]{This is an example for title footnote coding.}

\author[1]{Mauricio Orbes-Arteaga}
% \cortext[cor1]{Corresponding author}
% \ead{henry.m.orbes_arteaga@kcl.ac.uk }  
\author[3]{Thomas Varsavsky}
\author[2,4]{ Lauge Sørensen}
\author[2,]{Mads Nielsen}
% \fntext[fn1]{This is author footnote for second author.}
\author[2,4]{Akshay Pai}
\author[1]{Sébastien Ourselin}
\author[1]{Marc Modat}
\author[1]{M Jorge Cardoso}

\affil[1]{ Biomedical Engineering and Imaging Sciences, King’s College London, London, UK}
\affil[2]{Cerebriu A/S, Copenhagen, Denmark}
\affil[3]{Department of Medical Physics and Biomedical Engineering, UCL, London, UK}
\affil[4]{Department of Computer Science, University of Copenhagen, Copenhagen, Denmark}
\maketitle

\begin{abstract}
The insertion of deep learning in medical image analysis had lead to the development of state-of-the art strategies in several applications such a disease classification, as well as abnormality detection and segmentation. However, even the most advanced methods require a huge and diverse amount of data to generalize. Because in realistic clinical scenarios, data acquisition and annotation is expensive, deep learning models trained on small and unrepresentative data tend to outperform when deployed in data that differs from the one used for training (e.g data from different scanners). In this work, we proposed a domain adaptation methodology to alleviate this problem in segmentation models. Our approach takes advantage of the properties of adversarial domain adaptation and consistency training to achieve more robust adaptation. Using two datasets with white matter hyperintensities (WMH) annotations, we demonstrated that the proposed method improves model generalization even in corner cases where individual strategies tend to fail. 

%%%%
\end{abstract}

%\linenumbers

%% Recommended structure
\section{Introduction}
% What is know
Domain adaptation has become an important area of research in medical image analysis. In applications such a segmentation or classification, domain adaptation aims to improve machine learning models' generalization making them more applicable in clinical practice. 

Machine learning models trained under empirical risk minimization tend to under-perform under distribution shift between training (source domain) and testing data (target domain). When analysing medical images such as magnetic resonance images, differences in the acquisition scanners and protocols can result in differences distribution shift that affect the performance of even most sophisticated deep learning models.

Deep domain adaptation is the research area that focuses in adaptation of CNN based models, by learning representations that are domain invariant but still  meaningful for the task at hand. Several research works have been proposed for Deep domain adaptation, a detailed review with focus in computer vision applications can be found in  \citep{wang2018deep}, also \citep{cheplygina2019not} discussed domain adaptation methods in medical image analysis. Both works provide a similar taxonomy for domain adaptation methods which grouped them into reconstruction based methods, adversarial based methods and discrepancy based methods. 

%About reconstruction methods. 
Reconstruction based methods rely on the idea that a model can be trained to accomplish both, an objective task (e.g segmentation/classification) and an auxiliary task such a image reconstruction which can be learned from unlabeled data. By doing this, the auxiliary task acts as a regulariser that promotes   inter domain representations learning and consequently improving the performance on the target domain set. A disadvantage of this methods is that usually the reconstruction of medical images is troublesome due to presence of abnormalities which also are diverse in morphology and localisation. Auxiliary tasks that are more complicated than the objective task usually lead to bad distribution of the feature representation and consequently low performance.
On the other hand,  adversarial and domain adaptation methods have shown to be more efficient and therefore more popular in domain adaptation tasks. One disadvantage of adversarial domain adaptation is the need for an auxiliary model (generator/discriminator), which adds complexity to the training hyper-parameter tuning.

%aboutadversarial based methods. (those are also fine tuning approaches)
In contrast, discrepancy based methods also referred to as fine-tuning approaches rely on a single model. The objective is to leverage available target domain data (labeled / unlabeled) to adjust model feature representation in such a way it is indistinguishable of the domain. The approaches vary in the way the fine-tuning is performed. For example, \citep{tzeng2014deep} explicitly minimize maxi-mun mean discrepancy (MMD) between representation of source (labeled) and target (unlabeled) domain. Newer approaches based on self-learning aim to match feature representation through pseudo labeling of target domain images . Consistency training approaches such as \cite{french2017self,perone2019unsupervised,xie2019unsupervised} minimize implicit domain discrepancy through a consistency loss between pseudo predictions in the unlabeled data.

\citep{french2017self} and \citep{perone2019unsupervised} use mean teacher approach where a teacher model (average of previous iteration models) generate predictions for noisy versions of the target images whereas a student model is encouraged to learn the same predictions through a consistency loss. \citep{xie2019unsupervised} get rid of the teacher/student approach and force a single model to be consistent to different perturbations of the input image. 
Although the approach was originally proposed for unsupervised data augmentation, 
one of the main contributions of that work was the exploration of different augmentation and their effect on model performance; more sophisticated augmentation strategies achieved higher performance which is also directly correlated to their performance in a supervised setting. This aligns with findings in \citep{french2017self} where stronger augmentations were needed to cope with larger shifts domain distributions. Those findings demonstrated that augmentations are responsible to a large extent for distribution matching.

Based on the above mentioned findings, the method proposed int  in this work lies on the intersection of adversarial and consistency training based domain adaptation. We hypothesized that a pre-alignment of the domain distributions can help consistency training to cope with scenarios where augmentation is not enough to match domain distributions. We adapt the  unsupervised data augmentation approach from \citep{xie2019unsupervised} to perform domain adaptation in white matter hyperintensities segmentation task. 

Thus, the contributions of this paper are:

\begin{itemize}
    \item We boost domain consistency training performance by using different augmentation strategies. We follow the criterion on \citep{xie2019unsupervised} to design augmentations that are realistic, provide inductive bias, and are  domain-specific. Specifically, we try spacial based transformations. Task-specific augmentations such a Bias-field and k-space, We go further and try consistency loss between two acquisitions from the same sample that can be considered as realistic augmentations.
    \item We introduce an adversarial loss as a feature space pre-alignment strategy. We proved this pre-alignment helps to guide consistency training in adaptation scenarios where augmentations can not cope with domain shift differences. We also showed that adding an adversarial loss helps to prevent the models to obtain inconsistent segmentation that can arise for using consistency training alone. 
    \item  We compared our proposed method against both adversarial learning and other state-of-the art self-learning approaches. We use several metrics of performance and use a ranking scheme to evaluate the final performance of each method. 
\end{itemize}

This work is an extension of our preliminary work \citep{orbes2019multi}.  We extend the proposed method to work outside of the paired-data setting. We include the result in additional data sets which hold data from different clinics. 

\section{Methods}
\subsection{Distribution shift and unsupervised domain adaptation }
When learning a segmentation model  it is usually assumed that training and test sets belong to the same domain distribution. 

During the training stage, images $x_s$ and their respective annotations $y_s$ drawn from a distribution $P_s(x_s,y_s)$ (source domain) are used to learn a mapping $f_s(x_s) = P_s(y|x)$ that approximate the optimal function $f'_s(x_s)$ that generates $y_s$. 

However, in the testing stage $f_s()$ tend to under-performs on data draw from a target domain with a distribution $P_t(x_t,y_t)$ which differs from the distribution of the source $(P_t \ne P_s)$. 

Real world applications face this distribution shift, which results from differences in acquisition protocols, scanner , and image properties (intensity, resolution, modality).

% When learning a segmentation model  it is usually assumed that data from  training and test sets belong to the same domain distribution. 
% During the training stage, a set of images and their respective annotation $x_s,y_s$ drawn from a distribution $P_s(x_s,y_s)$ (source domain) is used to learn a mapping $f_s(x_s) = P_s(y|x)$ that approximate the optimal function $f'_s(x_s)$ that generates $y_s$. 
% However, in the testing stage $f's$ tend to under-performs when evaluated on data draw from a target domain with a distribution $P_t(x_t,y_t)$ which differs from the distribution of the source $(P_t \ne P_s)$. 
% Real world applications experience this distribution shift, which come from differences in acquisition protocols, scanners used, and image properties (intensity, resolution,modality).

Domain adaptation aims to learn a representation that can perform well under distribution shift. Adaptation can be supervised, where a small set of annotated images from the target distribution is used to  fine-tune a model initially pre-trained on a source domain set. The applicability of supervised approaches is however limited; Assuming access to annotated data for every target set is unreliable in a clinical scenario. Unsupervised domain adaptation overcomes that limitation by leveraging unlabeled data from the target domain. Formally, given the annotated set $S=(x_s,y_s)$ (source domain) and the unlabeled set  $T=(x_t)$ (target domain), the aim is to use $T$ to aid the learning of a function $f_a()$ that works well in both $P_s$ and $P_t$ distributions.

For efficient leveraging of unlabeled images, we proposed a domain adaptation methodology that combines augmentation-powered consistency training and adversarial learning. Although, consistency training is simple and achieved outstanding performance for domain adaptation \citep{french2017self,tarvainen2017mean}. Still It lacks robustness on situations with a high discrepancy between source and target distributions. In those situations, a pre-alignment of domain distributions is needed, a task in which adversarial learning is outstanding  \citep{kamnitsas2017efficient,ganin2016domain}.

\subsection{Consistency training:} 

Consistency training was originally proposed for semi-supervised learning as a way of generate predictions on unlabeled data. \citep{miyato2018virtual}. In the unsupervised domain adaptation setting, a consistency regularization term $\mathcal{L}_{consistency} $ is used on target domain samples (unlabeled) to force the model predictions to be invariant to different levels of noise in the input. 

% Consistency training demonstrated to be successful leveraging unlabeled data on semi-supervised learning settings. For unsupervised domain adaptation, a consistency regularization term $(\mathscr{L}_{consistency})$ is applied on target domain samples (unlabeled)  to force the model predictions to be invariant to different levels of noise in the input. 

Formally,  given  an image $x$ and its augmented counterpart $\hat{x}$ which is drawn from $q(x|\hat{x})$ where $q$ is an augmentation function. As proposed in \cite{xie2019unsupervised}, model consistency  is achieved  by minimizing the the Kullback-leibler divergence $\mathscr{D}( P(y|x) | P(y|\hat{x}))$. Where probability distributions $ P(y|x)$ and $P(y| \hat{x})$ are obtained from evaluating the model $f$ on $x$ and $\hat{x}$ respectively.
    
In this paper however, we replaced Kullback-leibler divergence by the dice loss \cite{milletari2016v}, which is better suited for the particularities of segmentation task and also because it proven to be effective to deal with class imbalance. 

The domain adaptation is then driven by a jointly optimization on source and target domains that computes: a supervised loss using only pairs $(x_s, y_s)$ from the source domain set $S$, and a consistency loss using only samples  $x_t$ from the target domain set $T$. The full objective loss is written as follows:
% Thus, a jointly optimization on source and target domains computes a supervised loss using only pairs $(x_s, y_s)$ from the source domain set $S$, and a consistency loss using only samples  $x_t$ from the target domain set $T$. The full objective loss for domain adaptation with consistency training can be written then as follows:
    \begin{equation}
        \mathcal{L}_{CT} = \mathcal{L}_{supervised} + \mathcal{L}_{consistency} =  dice(\hat{y}_s, y_s) + dice(\hat{y}_t, \hat{y}_{\hat{t}}) 
    \label{consistency-training}
    \end{equation}
Where $y_s$ is the ground truth, $\hat{y}_s$ is the prediction on source domain samples, $\hat{y}_t$ the prediction on the target domain samples and $\hat{y}_{\hat{t}}$ is the prediction on its augmented counterpart.

\subsection{Reinforcing consistency training through data augmentations:} 
    
 As shown in \citep{xie2019unsupervised},  the quality and properties of the function $q$ play an important role in consistency training. More natural and diverse augmented images will result in a performance gain; They make the model more robust to a wider range of variations beyond the ones seen in the training images.  

As suggested by \citep{xie2019unsupervised} we aim to achieve three properties on augmented samples when design augmentation functions: \textit{i)} samples should be \textbf{valid and realistic}, meaning that the original and augmented samples must share the same relative labels to safely encourage consistency in their predictions --in the context of MR-images, morphology should be preserved. \textit{ii)} Samples should be \textbf{diverse} enough in order to increase the model robustness to expected variability during the testing stage. \textit{iii)} Samples should provide  \textbf{targeted inductive biases} to help to capture differences in the domain knowledge. 

To cope with the above mentioned requirements, we proposed the following transformations:
    
\begin{itemize}
    \item \textbf{Geometric augmentation:} transforming the spacial relations in the images leads to generating a more diverse set of samples which also improves sample efficiency. Therefore, encouraging consistency to geometric augmentations can significantly improve robustness to such geometrical variations. We sample independently geometric augmentations which include, random rotations (all axis ranging from -10 to 10 degrees), random shears ($[0.5,0.5]$), and random scaling ($[0.75,1.5]$) and combined them as one affine transform.  We discard non-linear deformations because they increase the complexity of implementation when performed online, and could produce not realistic results. Geometric affine transformations are simpler to implement online, and they are safer to encourage consistency. 
    \item \textbf{MRI-specific augmentation: } Common artefacts caused during the acquisition process of MRIs could produce errors in segmentation models. Therefore, augmenting the data with synthetically generated artefacts will increase the image appearance variability and improve the robustness to those anomalies commonly observed in MRIs. We apply k-space motion artefact augmentation as described in~\cite{shaw2019mri} and bias field augmentation as implemented in~\cite{gibson2017niftynet}. 
    \item \textbf{Real:} We consider the case where we have access to different separate acquisition of the same subject (e.g 2D and 3D acquisitions). 
    If the samples are paired (one-to-one spatial correspondence and share the same labeling) they can be considered as realistic augmented versions of each other. Given each sequence belongs to a different domain distribution, encouraging consistency on their predictions makes the model robust to the distribution shift of their domains%--This can be also seen as a multi-domain adaptation approach, where we get the adaptation from one source to multiple  target domains. 
    . Taking the sequences as they are makes for a discrete augmentation function with discontinuous jumps.  In order to encourage continuity  we also apply to each acquisition a combination of both, geometric and MRI specific augmentations.
\end{itemize} 

\subsection{Aligning domain distributions:}

As pointed by \citep{french2017self} and \citep{miyato2018virtual} consistency training (which belong to the family of self-ensembling methods) work by label propagation from source to target domain. Therefore, on extreme cases where the domains are considerable different, the augmentation function could be insufficient to align both distributions. In those situations, consistency training can lead to under-performance or even segmentation failure. As we will show in one of our experiments, our model was able to identify the domain and using this information to inform the solution: An accurate predictions on the source domain, and a trivial solution on the target domain --a foreground segmentation mask for both $\hat{y}_u$ and $\hat{y}_{\hat{u}}$ that are sufficiently similar to erroneously inform high consistency.
% .This is an undesirable behavior because which resulted in an accurate solution for the source domain while producing trivial one on the target domain (in this case, a foreground segmentation mask). So $\hat{y}_u$ and $\hat{y}_{u}$ were sufficiently consistently to erroneously inform high consistency. 
Designing  customised functions for each adaptation case is unpractical and evaluating its alignment efficiency is ambiguous (there is not ground-truth labels). 

To tackle this problem, we incorporate an adversarial loss to align the domain distributions. Similar to \citep{kamnitsas2017unsupervised} we aim to find a feature representation $h$ that is domain invariant. In this work,  $h$ is obtained from a set of activation maps produced by $f$ (a neural network) when evaluating an input sample. Note that when $f$ is trained in one domain only, it will produce $p(h_s) \neq p(h_t)$ (for a given samples $x_s$ and $x_t$ respectively)  meaning both distributions are different. 

We use a discriminator network $d_\Omega$, which takes $h$ as input and produces a domain prediction $\hat{d}$. The loss that optimize the discriminator is: 
\begin{equation}
    \mathcal{L}_{adversarial} = \sum_{i=1}^n \mathcal{L}^{i}_{ce}(d_i, \hat{d_i})
    \label{adversarial}
\end{equation} 

Where, $\mathcal{L}^{i}_{ce}$ is the multi-class cross entropy loss, $d$ is a one-hot encoded vector of the domain label and $\hat{d}$ is the model's domain prediction as in \citep{schoenauer2019multi},  $n$ is the number of domains, and enables our framework to generalize to multiple domains.
% \item "classification accuracy serves as a discrepancy measure that that eliminates distribution shift (learning invariant features)" 

The classification accuracy informs how domain specific is the feature representation $h$. Thus, to get an invariant representation,  $d_\Omega$ and $f$ are  simultaneously trained to maximize $\mathcal{L}_{adversarial}$ while minimize $\mathcal{L}_{CT}$. 

\begin{figure*}[t!]
    \centering
    \includegraphics[scale=0.5]{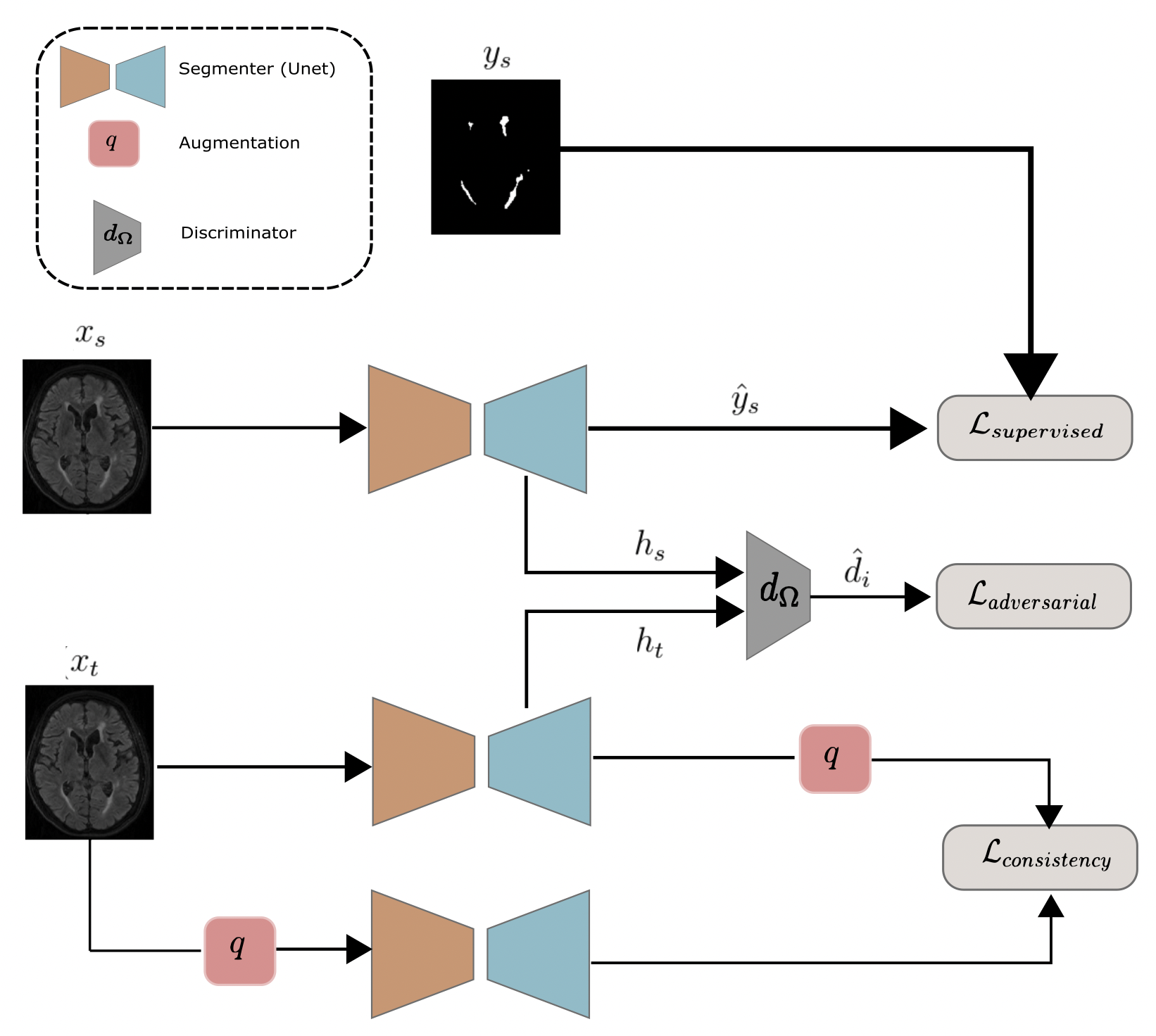}
    \caption{
    Diagram of proposed method. At training time, $x_u$, $x_l$ and $y_l$ are supplied to the network. $x_u$ is an image from the unlabeled target domain  and $\hat{x}_u$ is the result of applying some augmentation function to $x_u$.
    A labeled image, $x_l$, is passed through the network, $f_\theta$ before combining with a label $y_l$ to form the segmentation loss, $\mathscr{L}_s$. The image representations are fed to a domain discriminator $d_\Omega$ which attempts to maximise the cross-entropy between predicted domain and actual domain,  $\mathscr{L}_{adv}$.  Finally, similarity is promoted between the network predictions on $x_u$ and $\hat{x}_u$ using $\mathscr{L}_{PC}$.
    }
    \label{fig:method_scheme}
\end{figure*}

The full optimization loss used in our domain adaptation framework is then given by:
\begin{equation}
     \mathcal{L}_{total} = \mathcal{L}_{supervised} + \alpha \mathcal{L}_{consistency} - \beta \mathcal{L}_{adversarial}
     \label{eq:total_loss}
\end{equation}
We use $\alpha$ and $\beta$ to weigh the contribution of the consistency and the adversarial losses in the optimization. This also makes easier the scheduling of the training process which is carried out in three phases: During the first stage $\alpha$ and $\beta$ are set to 0, so the segmentation is only trained in a supervised fashion -in this stage, the network learn meaningful representations but they are source-specific. During the second stage, $\beta$ is set to one in order to learn invariant to domain feature representations. Finally, once the domains are aligned, it is safe to activate the consistency loss so $\beta$ is set to one, so all components in \Cref{eq:total_loss} contribute to the optimization. \Cref{fig:method_scheme}, shows the scheme of our domain adaptation framework. 

\section{Experiments and results}
We developed experiments to assess our method and the influence of each of its components (adversarial learning, augmentation, and consistency training).  Because the evaluation of realistic augmentations requires paired data (one-to-one pixel correspondences) we divided the experiments according to the available databases separately. 

\subsection{Evaluation on unpaired data} \label{sec:evaluation_unpaired}
\subsubsection{Task and data}
\noindent
In this experiment, we aim to assess our method in a domain adaptation set-up that is commonly seen in clinical practice: One or multiple clinics provide annotated images (source domain) for development, whereas, a different clinic that uses different acquisition protocols provides unlabeled images (target domain) for deployment. To simulate this scenario we use the White Matter Hyperintensity Segmentation Challenge Dataset \citep{kuijf2019standardized}, which holds MR brain images (T1-w and FLAIR  pairs ) for 60 subjects from three different clinics. For each pair, it also provides manual annotations of WMH. As pre-processing steps, T1-weighted images are registered to FLAIR. The images were also bias field corrected using SPM12. \Cref{WMH_dataset} show additional details for each clinic. 

% TODO: in our experiments we only use FLAIR as input for our models, as this was the reference for Clinicians manual labeling. 

\begin{table*}[t!]
    \caption{Summary of data characteristics in the WMH challenge database}
    \centering
    \begin{tabular}{lc@{\quad}c@{\quad}c@{\quad}cl}
    \hline\noalign{\smallskip}
     Clinic & Scanner Name & Voxel Size($m^3$) & Size & $\#$ scans\\
     \noalign{\smallskip}
\hline
\noalign{\smallskip}
     Utrech   &  3T Philips Achieva & \small{$ 0.96 \times 0.95 \times 3.00 $} & \small{$240 \times 240 \times 48$} & 20 \\
     Singapore &  3T Siemens TrioTim &  \small{$1.00 \times 1.00 \times 3.00 $} & \small{$252 \times 232 \times 48$} & 20 \\
     Amsterdam      &  3T GE Signa HDxt   &  \small{$ 1.20 \times 0.98 \times 3.00 $} & \small{$132 \times 256 \times 83$} & 20 \\ 
     \noalign{\smallskip}
\hline
     \end{tabular}
     \label{WMH_dataset}
\end{table*}
%maybe this goes after everything before results or could be also in the settings. 
\subsubsection{Implementation details} \label{sec:implementation_unpaired}
\textbf{Data configuration:}  we split the source domain data into training (50 $\%$), validation (25$\%$), testing (25$\%$) and the target domain into training (50 $\%$), and testing (50 $\%$). We feed our models with 2D slices of size  $256 \times 256$  extracted from FLAIR sequences. Note, we don't use T1-W images, as they were not considered for manual segmentation.
% TODO: give an explanation if necessary for why no introduce T1-w. 
% Network architectures. 
\textbf{Model Architectures:} The segmentation network uses a U-Net architecture \cite{ronneberger2015u} with a depth of 4 and a maximum number of filters of 256 at the deepest layer which also uses dropout. We use batch normalization and ReLu as activation function. The discriminator architecture is composed of four 2D convolutional layers with a kernel size of $3 \times 3$ and stride of 2 followed by batch normalisation, and leaky ReLu activation. The number of output channels is 4 to begin with and doubles at each layer to a total of 32. Finally, there are three fully connected layers with outputs sizes of 64, 32, and 2 which use ReLu activation and dropout (p$=$0.5).

\textbf{Model training:} Training of our proposed method occurs in three stages:
% TODO: Write in terms of iterations
% Training steps:
\begin{itemize}
    \item \textbf{Supervised training (only in source): } The model training starts by minimizing $\mathcal{L}_{seg}$  using only image pairs ($x_s,y_s$) from the source domain. We use the Adam optimizer with learning rate $10^{-3}$ and learning rate decaying schedule with $\gamma=0.1$ where $\gamma$ where, gamma is a multiplicative factor of learning rate decay, at epochs 30 and 350. We use the validation set from the source domain to monitor the performance, thus we select the model weights which achieve the highest dice score in the validation. Note, at this point, the segmenter network achieved a baseline of performance (No adaptation), and is used as a base to continue the next stages of learning.
    % TODO: explain learning rates are initialized again.
    \item \textbf{Adversarial training:} keeping the weights of the base model frozen,  we compute the target representation for source and target images and use them to train the discriminator for 30 epochs. Once the discriminator has achieved a desirable high accuracy, then we unfroze the weights on the segmenter and both networks are trained simultaneously by minimizing $\mathcal{L}_{supervised} - \beta \mathcal{L}_{adversarial}$, with ($\beta =0.3$). Training continues for 100 epochs until the discriminator is fooled (accuracy close to $50\%$), indicating the feature representation is invariant to the input domain. 
    \item \textbf{Consistency training:} Finally, we compute consistency training loss using a different batch of images $x_t$ from the target domain, at this point the whole framework is trained  by minimizing \Cref{eq:total_loss}. Training continues for 300 epochs, with learning rater decay at epochs 200 and 250. 
% TODO: Make a figure which explain the training processes. 
\end{itemize}

% Note in our experiment we only use FLAIR images, and show here the partitions. Maybe, shows an scheme about partitions here. 
% \subsubsection{implementation details}
% \begin{itemize}
%     \item Images description.
%     \item Segmentation network. 
%     \item Phase one on source.
%     \item Adam optimizer.
%     \item validation used for early stopping. 
%     \item All adaptation models were trained with the Initial weights using the baseline model. 
%     \item --
%     \item Parameters $\alpha,\beta$. 
%     \item For the domain discriminator initial settings
% \end{itemize}

% Implementation details from reuben.
%- Stochastic gradient descent, on loss which summs of seg and adaptation.
%- Batch extraction and how loss are computed. (how the pairs fit  inthe losses. 
% - domain adatpation via AUGMENTATION, explain about interdomain fature discrepance. 
% - domain adaptation via ADVERSARIAL,  explain steps and scheduling 
% It seems he does not combined both.

% Implementation details. 
% pytorch, layer initialization. , other parameters, discriminator,
% Cross validation, and partition sets
% Augmentation settings. 
% Optimizer
% Learning rates
% Probabilistic version of the jaccar loss function. to give same weight to segmentation and tissue segmentation (classes)
\subsubsection{benchmarks for comparison}
To evaluate our proposal we consider differ models for comparison:
\begin{itemize}
\item \textbf{No Adaptation:} We train a model  without adaptation using only annotated data from the source domain. This is equivalent to the base model on the first stage of our proposal. We perform inference in the test set from the target domain data. 
\item \textbf{Adversarial domain adaptation (Adversarial):} We perform adversarial domain adaptation following the guidelines in \citep{kamnitsas2017unsupervised}. Note, this is equivalent to the second stage of our method. We use the labeled training data from the source domain together with unlabeled training data from the target domain to minimize the loss function $\mathcal{L}_{ADA}=\mathcal{L}_{supervised}+ \mathcal{L}_{adversarial}$. 
\item \textbf{Mean teacher :} We train The Mean-Teacher algorithm for segmentation as proposed in \citep{perone2019unsupervised}. For a fair comparison, we complemented the mean Teacher with the proposed Geometric and MRI-specific augmentations. Consistency is trained with the target training set. 
\item \textbf{Training consistency with adversarial learning (TC $\&$ Adversarial ):} We train our proposed model using a combination of geometric and MRI-specific augmentations.
\item \textbf{Supervised:} We train the model using the training labeled images from both source and target domains. This method provides an upper-bound of accuracy as is expected higher performance when introducing target domain knowledge.
\end{itemize}

\subsubsection{Assessment:} For each method we compared their predicted 3D volumes segmentation against the ground truth labels --3D volumes results from concatenating 2D slice predictions--. As similarity measures  we use, Dice overlap, Hausdorff distance (HD-95), Recall, and ratio volume difference (VD) as described in \citep{kuijf2019standardized}." We follow the method described in Medical Decathlon Challenge \footnote{http://medicaldecathlon.com/files/MSD-Ranking-scheme.pdf} , to provide a single rank score comparing all methods. Note, this ranking provides a per-metric non-parametric statistical significance model.
% TODO: add a better description of significance ranking. 
\subsubsection{Results:}
% COMMENT: only one technique for comparison is not good enough, multiple techniques provides a better overview.
For each cross clinical experiment, we show in \cref{tab:allresults} the average scores for each metric and the ranking score (p $<$ 0.01 paired Wilcoxon test) for each method. In addition, we show in \cref{tab:rankall} the average ranking score computed across all experiments. Overall, combining consistency training with adversarial learning outperformed individual implementations for consistency training and adversarial methods. The advantage of the proposed combination is notorious in the experiment \textit{clinic1 $\rightarrow$ clinic3} and \textit{clinic2 $\rightarrow$ clinic1 }  
where differences between source and target domains can not be matched through augmentations only, so a feature representation alignment is needed. In addition, the proposed training consistency outperformed the mean teacher, showing the advantages of the proposed optimization strategy.
\begin{table*}[t!]
    \centering
\begin{tabular}{llllll}
\toprule
                                  Method &            Dice &     HD95 &       VD &            Recall     \\
\midrule
\textbf{Clinic1  $\rightarrow$ clinic2} & & & & & \\
No adaptation & 0.6761(0.14) & 7.5719(8.16) & 0.3744(0.10) & 0.5514(0.13) & 4.50 \\
Mean Teacher & 0.6834(0.16) & 7.7939(9.28) & \textbf{0.2784}(0.14) & 0.5835(0.13) & 3.67 \\
Adversarial & 0.7141(0.13) & 5.5040(7.66) & 0.3514(0.08) & 0.5899(0.12) & 4.50 \\
TC & 0.7392(0.14) & 6.8165(9.60) & 0.3197(0.10) & 0.6260(0.14) & 3.83 \\
\textbf{TC $\&$ Adversarial} & \textbf{0.7572}(0.10) & \textbf{3.8137}(3.46) & 0.3262(0.10) & \textbf{0.6373}(0.11) & \textbf{3.50} \\
Supervised & 0.8220(0.09) & 3.0092(5.16) & 0.1601(0.11) & 0.7599(0.11) & 1.00 \\
\midrule
\textbf{Clinic1  $\rightarrow$ clinic3} & & & & & \\
No adaptation & 0.6677(0.13) & 10.4082(8.04) & 0.2460(0.15) & 0.6246(0.16) & 4.83 \\
Mean Teacher & 0.6758(0.12) & 8.9834(6.28) & 0.2383(0.16) & 0.6534(0.15) & 4.83 \\
\textbf{Adversarial} & \textbf{0.7241}(0.11) & 6.0232(5.31) & \textbf{0.2083}(0.15) & \textbf{0.6635}(0.16) & \textbf{2.17} \\
TC & 0.7165(0.11) & 5.7597(4.08) & 0.2579(0.18) & 0.6314(0.15) & 4.33 \\
TC $\&$ Adversarial & 0.7230(0.12) & \textbf{5.2384}(4.32) & 0.2536(0.17) & 0.6419(0.16) & 2.83 \\
Supervised & 0.7498(0.09) & 3.8950(2.61) & 0.2222(0.16) & 0.6716(0.13) & 2.00 \\
\midrule
\textbf{Clinic2  $\rightarrow$ clinic1} & & & & & \\
No adaptation & 0.5927(0.24) & 9.5052(6.74) & 1.6038(2.73) & 0.8007(0.17) & 4.33 \\
Mean Teacher & 0.5890(0.22) & 8.7965(5.96) & 1.4889(1.95) & \textbf{0.8447}(0.13) & 4.33 \\
\textbf{Adversarial} & 0.6553(0.12) & \textbf{7.7815}(4.75) & \textbf{0.4723}(0.60) & 0.7237(0.17) & \textbf{3.50} \\
\textbf{TC} & 0.6199(0.24) & 8.0445(6.42) & 1.1809(1.80) & 0.8055(0.16) & \textbf{3.50} \\
\textbf{TC $\&$ Adversarial} & \textbf{0.6568}(0.19) & 8.2080(5.43) & 0.5972(0.73) & 0.7574(0.15) & \textbf{3.50} \\
Supervised & 0.6588(0.22) & 7.9915(5.97) & 0.6541(0.90) & 0.7688(0.15) & 1.83 \\
\midrule
\textbf{Clinic2  $\rightarrow$ clinic3} & & & & & \\
No adaptation & 0.6867(0.12) & 8.6575(8.95) & 0.2184(0.16) & 0.6493(0.16) & 4.50 \\
Mean Teacher & 0.6931(0.12) & 8.6861(6.20) & 0.2220(0.21) & \textbf{0.7427}(0.15) & 4.50 \\
Adversarial & 0.6759(0.12) & 5.7488(4.59) & 0.1968(0.15) & 0.6418(0.17) & 4.50 \\
\textbf{TC} & \textbf{0.7339}(0.10) & 5.0247(4.70) & \textbf{0.1697}(0.14) & 0.7074(0.16) & \textbf{2.33} \\
TC $\&$ Adversarial & 0.7205(0.10) & \textbf{4.7457}(4.67) & 0.1713(0.11) & 0.7276(0.15) & 2.50 \\
Supervised & 0.7402(0.09) & 4.8522(4.74) & 0.1911(0.15) & 0.6752(0.13) & 2.67 \\
\midrule
\textbf{Clinic3  $\rightarrow$ clinic1} & & & & & \\
\textbf{No adaptation} & 0.5856(0.26) & 10.8941(6.55) & 1.3213(3.03) & \textbf{0.7193}(0.15) & \textbf{4.00} \\
\textbf{Mean Teacher} & \textbf{0.6230}(0.23) & 10.1573(5.62) & 1.2807(2.79) & 0.7104(0.13) & \textbf{4.00} \\
\textbf{Adversarial} & 0.6177(0.23) & \textbf{9.0018}(6.14) & \textbf{1.0783}(2.48) & 0.6414(0.14) & \textbf{4.00} \\
\textbf{TC} & 0.6186(0.24) & 10.5676(6.00) & 1.2647(2.88) & 0.6898(0.15) & \textbf{4.00} \\
\textbf{TC $\&$ Adversarial} & 0.6217(0.23) & 9.1593(5.94) & 1.0821(2.48) & 0.6426(0.14) & \textbf{4.00} \\
Supervised & 0.6799(0.19) & 7.3986(5.90) & 0.3165(0.33) & 0.7180(0.15) & 1.00 \\
\midrule
\textbf{Clinic3  $\rightarrow$ clinic2} & & & & & \\
No adaptation & 0.7236(0.10) & 5.6932(7.41) & 0.2963(0.08) & 0.6288(0.09) & 5.17 \\
Mean Teacher & 0.7337(0.12) & 7.0330(8.98) & 0.2622(0.08) & 0.6610(0.10) & 3.67 \\
Adversarial & 0.6455(0.10) & 7.1411(8.65) & 0.3832(0.19) & 0.5333(0.11) & 5.67 \\
\textbf{TC} & \textbf{0.7690}(0.11) & \textbf{5.1755}(7.74) & \textbf{0.2127}(0.10) & \textbf{0.7112}(0.10) & \textbf{2.33} \\
TC $\&$ Adversarial & 0.7455(0.09) & 6.9319(9.00) & 0.2591(0.10) & 0.6632(0.07) & 3.17 \\
Supervised & 0.8286(0.07) & 2.4007(2.08) & 0.1540(0.08) & 0.7841(0.10) & 1.00 \\
\bottomrule
\end{tabular}
    \caption{ We report Dice, HD95, volume difference(VD) and Recall, that were computed between our prediction and the ground truth labels. A significance rank is calculated across all metrics.Results are reported with the format median (IQR) in percentages for all metrics except the HD95 in mm. Best results are in bold when significantly better than all others ($p < 0.001$ paired Wilconxon test) }
    \label{tab:allresults}
\end{table*}

\begin{table*}[t!]
    \centering
\begin{tabular}{lr}
\toprule
Method&  Average Ranking \\
\midrule
Supervised          &         1.583333 \\
TC  $\&$  Adversarial &         3.250000 \\
TC                  &         3.388889 \\
Adversarial         &         4.055556 \\
Mean Teacher        &         4.166667 \\
No adaptation       &         4.555556 \\
\bottomrule
\end{tabular}
    \caption{Average Significance ranking for each method, the mean is computed from the six cross clinical experiments}
    \label{tab:rankall}
\end{table*}

\subsection{Evaluation on paired data} \label{sec:evaluation_paired}
\subsubsection{Task and data:}
The objective of this experiment is to evaluate the effects of incorporating realistic augmentation in consistency training. To this end, we use a set of 72 MRIs from a sub-study within the Pharmacokinetic and Clinical Observations in PeoPle over fiftY (POPPY). In this study, for each subject, two different FLARI sequences were acquired during the same MR session on a Philips 3T scanner. The in-plane FLAIR was an axial acquisition with 3mm slice thickness and 1 $mm^2$ planar resolution (Repetition time (TR) 8000 m, Inversion time (TI) 2400 ms and echo time (TE) 125 ms) while the volumetric FLAIR was of resolution $1.04 \times 1.04 \times 0.56$ $mm^3$ (TR=8000ms TI=1650ms TE=282ms).  Both images were rigidly co-registered to the 1mm3T1 sequence acquired during the session. All individuals were male with a mean age of 59.1±6.9 yrs, including HIV-positive subjects and population-matched controls.

\subsubsection{Implementation details}\label{sec:implementation_paired}
\textbf{Data configuration: } We use as a source domain the entire dataset from the WMH segmentation challenge and the POPPY data-set as the target domain. We split both source and target data into training:validation:test, with assignments of 40:10:10 subjects for the source domain, and  38:15:20, for the target domain. 

\noindent
\textbf{Model architectures:} For the segmenter, we use the same model configuration presented in \Cref{sec:implementation_unpaired}. Here, we applied the same transformation to both, the 2D and 3D sequences, therefore, their outputs are directly comparable and consistency is computed without any extra trasformation. 

\noindent
\textbf{Model Training:} We keep supervised and consistent training stages as in \ref{sec:implementation_unpaired}, whereas, we slightly modify the adversarial training to take a batch of target domain images that contains the same amount of 2D and 3D sequences. Note, 2D and 3D sequences get the same label. We also found that high values of $\alpha$ produce degenerate solutions, such as predictions of background or other structures different that lesions. Since scheduling a slowly increasing $\alpha$ did not help, $\alpha$ was fixed at 0.2 in all experiments.\looseness=-1. Thus, in this scenario, the adversarial component plays an important role.

\subsubsection{Benchmarks for comparison and assessment:}

In this scenario, we assess the performance for the same methods presented in \ref{sec:evaluation_unpaired}. In addition, we observe the effect of applying augmentations on training consistency, adversarial, and no adaptation methods, so those methods were implemented with and without applying augmentations. Note, that training consistency without augmentations (TC No-Aug) is equivalent to training consistency with real augmentations only.

As the first metric of consistency, we compute the Dice score overlap between the two volumes. However, high dice agreement may arise without predicting lesions, for instance with the segmentation of the foreground or of another anatomical structure.  Such degenerate solutions can indeed occur as the consistency term in the loss can be minimized for any consistent prediction between volumes. 
As there is no lesion segmentation for the POPPY dataset, we use the known association between age and white matter hyperintensity load reported for this dataset~\citep{haddow2019magnetic} as surrogate evaluation that the segmented elements are lesions. The effect size is a useful metric for determining whether the lesion loads predicted by the various models agree with the reported literature. For the eight compared models, the effect size ranged from 1.2-fold to 1.5-fold increase in lesion load normalized by total intracranial volume per decade. This compares well with the reported effect size on the POPPY dataset of 1.4-fold with a 95\textsuperscript{th} confidence interval of $\left[1.0 ; 2.0\right]$. Predictions from in-plane POPPY and volumetric POPPY were compared using the dice overlap, the  95\textsuperscript{th} percentile Hausdorff distance measured in mm (HD95), the recall (or sensitivity), the ratio of difference in volume between the two predictions (VD) as in \citep{kuijf2019standardized}. 

\subsubsection{Results}
\Cref{tab:quantitative_paired} shows the median and interquartile range ordered according to the average significance ranking. Overall, results show a similar trend to the one presented in \ref{sec:evaluation_unpaired}, with the proposed training consistency methods outperforming mean teacher and adversarial implementations. In addition, we observed that adding geometric and MR-specific augmentation to the realistic augmentation improves the segmentation performance. One particularity was that adversarial training showed the lowest performance, also in this case the augmentation did not have a positive effect.  
\begin{table*}[t!]
  \centering
  \caption{Performance of different methods on the target  (POPPY) and the source domain (MICCAI 2017 WMH Challenge). We report the dice between our models' predictions and the ground truth annotations in the source domain as well as the HD95. The evaluation on target domains is done with the Dice, the HD95, the volume difference (VD) and the recall. A significative rank measure is calculated across all metrics. Results are reported with the format median (IQR) in percentages for all metrics except the HD95 in mm. Best results are in bold andunderlined when significantly better than all others (p$<$0.05 paired Wilcoxon tests).}
  \label{tab:quantitative_paired}
  \centering
    \begin{tabular}{cccccccc}
    \toprule
          & \multicolumn{4}{c}{POPPY}     & \multicolumn{2}{c}{MICCAI} &  \\
          & Dice  & HD    & VD    & Recall & Dice  & HD    & \multicolumn{1}{l}{Rank } \\
          \midrule
    TC $\&$ adversarial     & \textbf{\underline{54.5} (10.6)} & 32.7 (9.8) & \textbf{\underline{15.2} (22.8)} & \textbf{\underline{52.4} (14.4)} & 81.4 (9.6) & 28.5 (8.6) & 2.5 \\
    TC   & 53.2 (15.1) & 39.2 (15.5) & 25.4 (15.6) & 43.5 (12.5) & 81.6 (15.5) & 18.6 (4.8) & 3.3 \\
    TC No-Aug    & 50.7 (17.0) & 35.1 (11.9) & 16.6 (21.4) & 43.6 (11.0) & 81.4 (22.6) & \textbf{17.2 (3.6)} & 3.4 \\
    Mean Teacher    & 48.6 (12.3) & 33.6 (14.8) & 33.7 (19.0) & 40.9 (5.0) & 80.0 (18.2) & 20.0 (7.3) & 4.3 \\
    No Adaptation     & 42.8 (14.6) & 34.9 (11.1) & 39.3 (22.3) & 33.5 (12.6) & 80.6 (14.8) & 17.8 (4.9) & 4.9 \\
    No adaptation - No Aug    & 43.0 (16.2) & 33.3 (15.1) & 40.3 (24.8) & 33.3 (14.8) & 81.1 (16.9) & 17.5 (3.3) & 5.6 \\
    Adversarial-No Aug  & 41.8 (15.4) & \textbf{32.6 (6.1)} & 25.2 (24.0) & 33.5 (12.7) & \textbf{82.5 (12.0)} & 17.6 (5.2) & 5.7 \\
    Adversarial & 41.4 (16.4) & 36.6 (9.0) & 38.0 (16.0) & 33.6 (13.9) & 81.9 (11.1) & 19.7 (11.0) & 6.3 \\
    \bottomrule
    \end{tabular}%
    
\end{table*}%

\begin{figure*}[t!]
    \centering
    \includegraphics[width=0.80\textwidth]{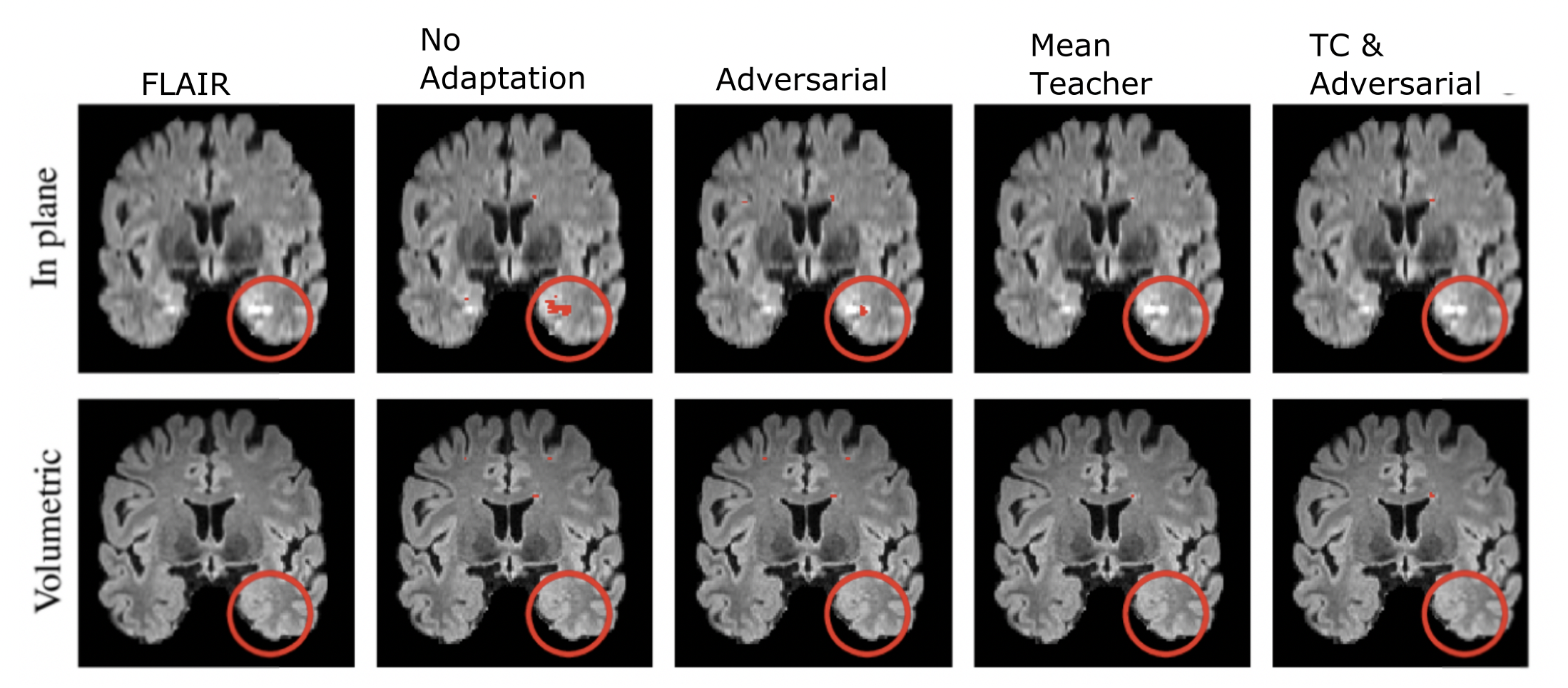}
    \caption{Qualitative results on a single slice from a single subject in the POPPY dataset. The top row shows a slice from the in-plane FLAIR acquisition whilst the bottom row shows a slice from the volumetric FLAIR acquisition. Each column shows a model's predictions on that row's image. This slice is used to highlight an example of an artefact (shown in the red circle) introduced by the in-plane acquisition. The baseline method introduces a false positive in this region whilst the domain adaptation methods perform better at ignoring it. Our approach shows the best in-plane to volumetric agreement.}
    \label{fig:qualitative}
\end{figure*}

\subsection{Assessing effect of  augmentations:} We evaluate the performance of training consistency when combined with different types of augmentations. We perform the cross-clinical experiments following the same configuration and data splinting as in  \cref{sec:evaluation_unpaired}. We compare consistency training (without adversarial learning) when combined with Geometric (TC-geo), MR-specific (TC-MR), and combination of all augmentations (TC-all). In addition, we perform consistency training without any augmentation (TC No-aug). 

\Cref{tab:resultsaug} Summarize the result for all metrics and the ranking score computed for each experiment. In addition, the average ranking score across the experiments is shown in \Cref{tab:rankaug}. In general consistency training is benefited from augmentation and combining different augmentation leads to the best performance.

\begin{table*}[t!]
\centering
\begin{tabular}{llllll}
\toprule
                                  Method &            Dice &     HD95 &       VD &            Recall &    Rank\\
\midrule
 \textbf{clinic1  $\rightarrow$ clinic2} &                 &                &               &                 &        \\
                     TC No-aug &    66.89  (13.0)   &    6.29  (9.82)  &   0.43  (0.09)  &   52.76  (12.38)  &  3.625 \\
                     TC Geo    &    70.6   (13.02)   &    \textbf{5.61  (7.6)}   &   0.37  (0.07)  &   57.82  (12.14)  &   2.75 \\
                  TC + MRI-aug &    73.74  (12.2)   &    5.67  (8.35)  &    0.34  (0.1)  &   61.67  (12.54)  &  2.125 \\
                  TC + all-aug &    \textbf{73.92  (13.58)}  &    6.82  (9.6)   &    \textbf{0.32  (0.1)}  &    \textbf{62.6  (13.51})  &   \textbf{ 1.5} \\
 \textbf{clinic1  $\rightarrow$ clinic3} &                 &                &               &                 &        \\
                     TC No-aug &   70.45  (11.04)  &    5.74 (3.44)  &   0.32  (0.17)  &    60.0  (14.45) &  3.375 \\
                        TC Geo &   \textbf{73.14  (9.47})  &    \textbf{4.82  (3.61})  &   \textbf{0.27  (0.16)}  &   \textbf{64.04  (13.62)}  &  \textbf{1.375} \\
                  TC + MRI-aug &   68.77  (12.19)  &    6.43  (4.07)  &   0.34  (0.17)  &   57.77  (15.24)  &  3.625 \\
                  TC + all-aug &   71.65  (11.01)  &    5.76  (4.08)  &   0.26  (0.18)  &   63.14  (14.83)  &  1.625 \\
 \textbf{clinic2  $\rightarrow$ clinic1} &                 &                &               &                 &        \\
                     TC No-aug &    62.3  (23.17)   &    8.51  (6.21)  &   1.28  (2.06)  &   \textbf{81.52  (14.87)}  &  2.625 \\
                        TC Geo &    61.03  (24.7)   &    8.09  (6.71)  &   1.37  (2.24)  &   80.38  (18.05)  &  3.125 \\
                  TC + MRI-aug &    \textbf{62.51  (23.14) } &    \textbf{7.93  (6.31)}  &   \textbf{1.12  (1.81)}  &   79.61  (15.07)  &      \textbf{2} \\
                  TC + all-aug &    61.99  (23.69)  &    8.04  (6.42)  &    1.18  (1.8)  &    80.55 (15.8)  &   2.25 \\
 \textbf{clinic2  $\rightarrow$ clinic3} &                 &                &               &                 &        \\
                     TC No-aug &   73.24  (10.18)  &    5.03  (4.79)  &   0.18  (0.14)  &   69.81  (14.97)  &  2.625 \\
                        TC Geo &    \textbf{74.94  (9.45})  &    \textbf{3.95  (3.86)}  &   0.18  (0.12)  &   \textbf{73.29  (15.04)}  &  \textbf{1.375} \\
                  TC + MRI-aug &   72.33  (10.99)  &    5.31  (5.18)  &    0.2  (0.16)  &   67.89  (15.85)  &  3.375 \\
                  TC + all-aug &   73.39  (10.49)  &     5.02  (4.7)  &   \textbf{0.17  (0.14)}  &   70.74  (15.57)  &  2.625 \\
 \textbf{clinic3  $\rightarrow$ clinic1} &                 &                &               &                 &        \\
                     TC No-aug &   61.71  (24.21)  &   12.69  (7.61)  &   1.43  (3.33)  &   69.85  (14.05)  &  2.875 \\
                        TC Geo &   61.74  (24.92)  &   13.04  (7.54)  &    1.52  (3.6)  &    \textbf{70.8  (15.06)}  &  2.375 \\
                  TC + MRI-aug &   \textbf{61.94  (22.97)}  &   \textbf{10.33  (6.42)}  &   \textbf{1.15  (2.49)}  &   68.99  (14.38)  &  2.375 \\
                  TC + all-aug &   61.86  (24.02)  &    10.57  (6.0)  &   1.26  (2.88)  &   68.98  (14.88)  &  2.375 \\
 \textbf{clinic3  $\rightarrow$ clinic2} &                 &                &               &                 &        \\
                     TC No-aug &   72.85  (11.81)  &    5.78  (8.04)  &   0.29  (0.06)  &     64.8  (9.28)  &      4 \\
                        TC Geo &   76.29  (11.44)  &     5.6  (8.11)  &   0.24  (0.08)  &    69.58  (9.96)  &   2.25 \\
                  TC + MRI-aug &    76.66 ( 8.91)  &    \textbf{4.55  (6.16)}  &   0.23  (0.08)  &    69.64  (9.66)  &   2.25 \\
                  TC + all-aug &    \textbf{76.9  (10.58)}  &    5.18  (7.74)  &    \textbf{0.21  (0.1)}  &    \textbf{71.12  (9.72)}  &    \textbf{1.5} \\
\bottomrule
\end{tabular}
\caption{Performance of different augmentation combinations for each cross-clinical setting, We report Dice, HD95, volume difference(VD) and Recall, that were computed between our prediction and the ground truth labels. A significance rank  is calculated across all metrics.
Results are reported with the format median (IQR) in percentages for all metrics except the HD95 in mm. Best results are in bold}
\label{tab:resultsaug}
\end{table*}

\begin{table*}[t!]
    \centering
\begin{tabular}{lr}
\toprule
              Method &      Rank \\
\midrule
    TC + all-aug &  1.979167 \\
        TC Geo &  2.208333 \\
          TC + MRI-aug &  2.625000 \\
 TC No-aug &  3.187500 \\
\bottomrule
\end{tabular}
\caption{Average Significance ranking computed for PC with different augmentations}
    \label{tab:rankaug}
\end{table*}

\section{Discussion}
We propose a methodology for robust domain adaptation which benefits from adversarial and consistency training strategies. Although we found that training consistency combined with robust augmentation improves model generalization, adversarial learning can boost their performance and robustness when augmentations are not sufficient enough to cover the gap between target and source domains. 

Specifically, the proposed training consistency method achieve higher performance when compared to individual implementations of adversarial and consistency training. It is worth nothing that our consistency training implementation achieved higher performance than  Mean Teacher \citep{perone2019unsupervised}, that improvement can come from the differences in their optimization strategies --By optimizing only one network we can achieve more smoother and continued updating of parameters compared to the moving average used on mean teacher.

Consistency training achieved overall better performance than its adversarial counterpart. However in those cases where domain distributions differ considerably, the adversarial domain adaptation  does a better job. This justify our idea of pre-aligning domains before apply consistency training. 

We found a more complex scenario on our paired experiments, augmentations were not enough to smooth differences in domain distributions so the model was able to produce domain specific solutions; a foreground segmentation for the target domain and wmhs segmentation for the source domain. In this case the role of the adversarial component comprises the elimination of the domain signal to avoid optimization be driven by trivial high consistency. Contrary to results on paired experiments, adversarial domain adaptation alone has a negative effect in segmentation performance, we hypothesize that depending on the adaptation problem the learning of a latent space invariant to domain  (as enforced in the adversarial approach) may cause an information loss detrimental to the segmentation task. 

% Augmentation
Augmentations plays and important role in consistency training optimization. We observed that the proposed augmentation strategies perform differently in each cross clinical experiment. We hypothesize the effectiveness for each augmentation depends on the distance of the source and target distributions, and the ability of each augmentation to bridge that gap. The overall higher performance (according to ranking score) when combining all the proposed augmentations could be explained because more variety of augmentations can be covered helping to smooth the domain differences. 

Although we advice to combine different augmentations to boost consistency training domain performance, one of the limitations of the proposed approach is not taking advantage of the domain distribution Knowledge to design
custom augmentations. Future work will focus in finding strategies for computation of domain distribution differences, that enable us to chose or select the right combination of augmentation functions. This strategies could also help to identify when adversarial learning is needed, so we can avoid any detrimental loss. 

\section{Acknowledgments}
This project has received funding from the EU H2020 under the Marie Sklodowska-Curie grant agreement No 721820.

% also \citep{cheplygina2019not} discussed domain adaptation methods in medical image analysis. Both works provide a similar taxonomy for domain adaptation methods which grouped them into reconstruction based methods, adversarial based methods and discrepancy based methods.

% \section{Discussion}
% % This should explore the significance of the results of the work, not repeat them. A combined Results and Discussion section is often appropriate. Avoid extensive citations and discussion of published literature.

% \section{Conclusions}
% % The main conclusions of the study may be presented in a short Conclusions section, which may stand alone or form a subsection of a Discussion or Results and Discussion section.
% We're cool

%%Harvard
\typeout{}

\bibliographystyle{unsrtnat}
\bibliography{refs}

\cleardoublepage
\appendix
%Appendices
%If there is more than one appendix, they should be identified as A, B, etc. Formulae and equations in appendices should be given separate numbering: Eq. (A.1), Eq. (A.2), etc.; in a subsequent appendix,
%Eq. (B.1) and so on. Similarly for tables and figures: Table A.1; Fig. A.1, etc.

%% main text

% \section*{Supplementary Material}

\end{document}